\documentclass[12pt]{iopart}
\newcommand{\gguide}

\usepackage{amssymb}
\usepackage{graphicx}
\usepackage{dcolumn}
\usepackage{bm}
\usepackage{color}

\begin{document}

\title{Electronic, Magnetic and Transport Properties of Graphene Ribbons Terminated by Nanotubes}

\author{M. A. Akhukov, Shengjun Yuan$^{1}$, A. Fasolino and M. I. Katsnelson}
% \author{Shengjun Yuan}
% \author{A. Fasolino}
% \author{M. I. Katsnelson}
\address{Radboud University of Nijmegen, Institute for
Molecules and Materials, Heijendaalseweg 135, 6525 AJ Nijmegen,
The Netherlands} 
\ead{$^{1}$s.yuan@science.ru.nl}

\date{\today}

\begin{abstract}

We study by density functional and large scale tight-binding transport calculations the electronic structure, magnetism and transport properties of the recently proposed graphene ribbons with edges rolled to form nanotubes. Edges with armchair nanotubes present magnetic moments localized either in the tube or the ribbon and metallic or half-metallic character, depending on the symmetry of the junction. These properties have potential for spin valve and spin filter devices with advantages over other proposed systems. Edges with zigzag nanotubes are either metallic or semiconducting without affecting the intrinsic mobility of the ribbon. By varying the type and size of the nanotubes and ribbons offers the possibility to tailor the magnetic and transport properties, making these systems very promising for applications. 

\end{abstract}

% \textbf{keywords: graphene, nanoribbon, nanotube, magnetism, spintronics} 

% \section{Introduction}

\begin{figure}[tbp]
\centering
\includegraphics[width=0.9\textwidth]{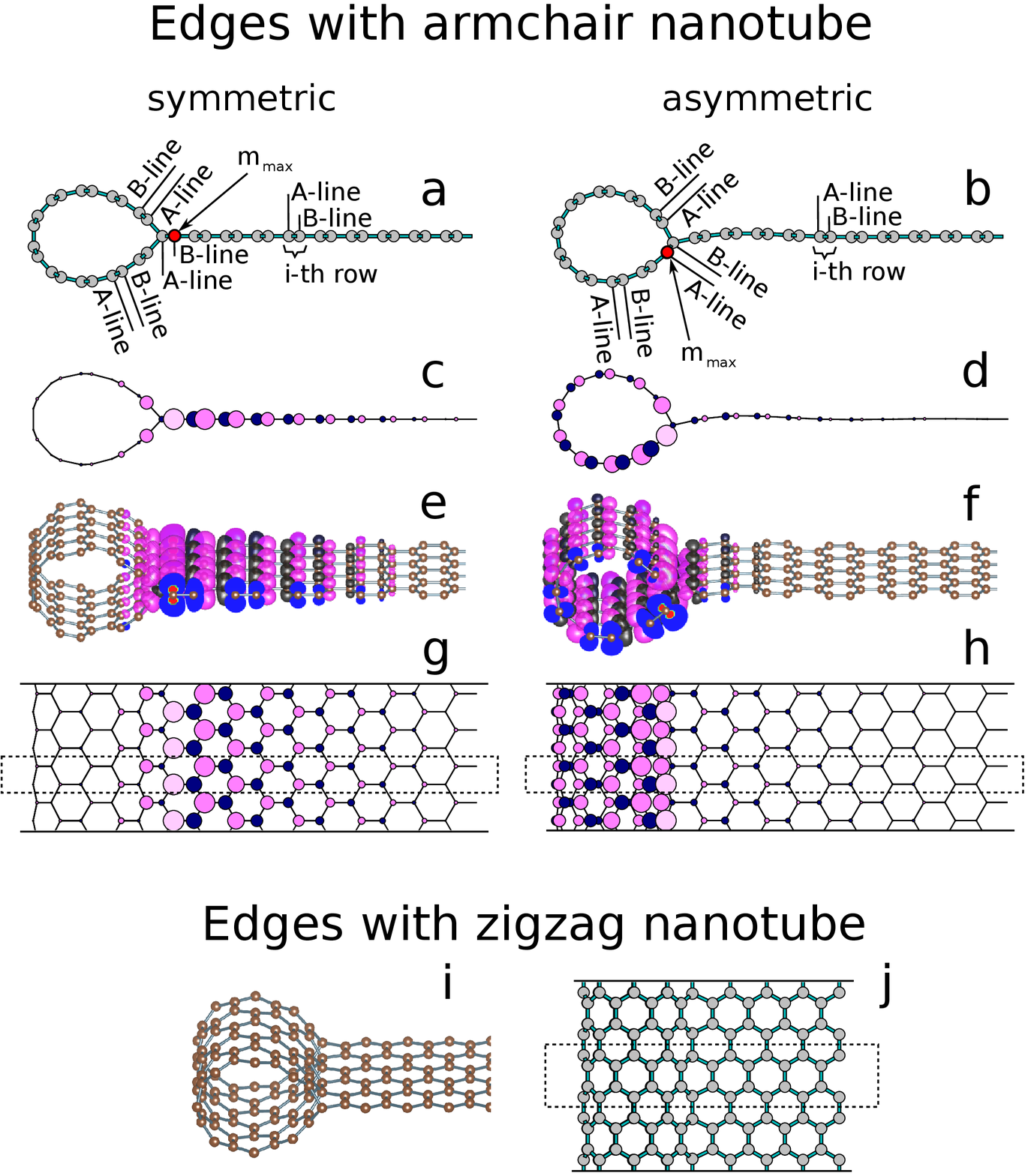}
\caption{(a-h) Structure and spin density for symmetric (left) and asymmetric (right)
AC nanotube terminated edges. (a,b) Atomic structure; (c,d) Side,
(e,f) 3D and (g,h) top view of schematic spin representation. The box with
dashed line in (g,h) indicates half of the unit cell in the DFT calculation.
For the symmetric case, the contribution to the magnetic moment from atoms
belonging to the A- and B- sublattices are $m_{A}$ = 0.958 $\protect\mu _{B}$
and $m_{B}$ = -0.217 $\protect\mu _{B}$ (per half unit cell), and for the 
asymmetric case $m_{A}$ = 1.048 $\protect\mu _{B}$ and $m_{B}$ = -0.300 $%
\protect\mu _{B}$. The maximum of the magnetic moment is located on the 
atom indicated by the arrow with the value $m_{max}$ = 0.379 $\protect\mu _{B}$
in both symmetric and asymmetric cases. (i,j) 3D and top view of AC nanoribbons terminated by ZZ nanotubes. The box with
dashed line in (j) indicates half of the unit cell in the DFT calculation.}
\label{fig:zzacnt}
\end{figure}

\begin{figure}[t]
\begin{center}
\mbox{
\includegraphics[width=0.8\textwidth]{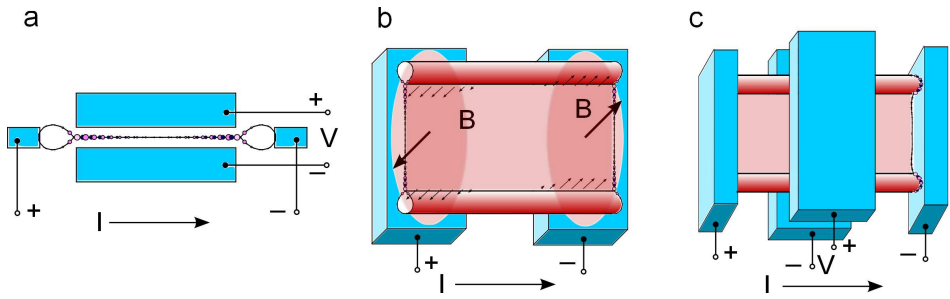}
}
\end{center}
\caption{Sketch of spintronics devices based on carbon nanoribbons terminated  by AC nanotubes. 
 (a) A spin valve based on the symmetric case: a gate can be used to switch  from the antiferromagnetically coupled state to the ferromagnetically coupled excited state, favouring spin transport from one nanotube to the other across the ribbon. (b)  For both symmetric and asymmetric case,  high magnetoresistance could be achieved by applying magnetic fields of different sign at the ends of the nanoribbons, as proposed in \cite{KimWY2008} for ZZ nanoribbons.
(c)  For the asymmetric case, a gate along the ribbon could be used to switch  between two half-metallic energy regions  to realize either a spin filter or a spin valve. 
 }
\label{app}
\end{figure}

The atomic structure of graphene edges is important for the determination
of the electronic and magnetic properties of graphene, especially for narrow
graphene nanoribbons\cite{Son2006,Han2007,Koskinen2008,Wassmann2008,Girit2009,LiuZ2009,Wimmer2010,Yazyev2010,Ostaay2011,JiaXT2011,Kunstmann2011,KatslesonBook}.
Recent theoretical work\cite{Ivanovskaya} on the stability of different graphene
edges structures has shown that graphene edges can fold back on
themselves and reconstruct as nanotubes, with low formation energy (see atomic structures in figure \ref{fig:zzacnt}). In this article, we show that, beside protecting the edges from contamination and reconstructions, 
nanotubes at the edges may lead to magnetism and are not detrimental for the electronic mobility despite the row of $sp^{3}$ hybridized  atoms at the ribbon-tube junction.  We study the electronic and magnetic 
properties of these systems by a combination of density functional theory (DFT) and 
large scale tight binding (TB) simulations of transport properties. Our calculations suggest that these systems could be used for a variety of applications that we sketch in figure \ref{app}. 

We consider systems formed by a nanoribbon terminated on both sides by the same armchair (AC) or zigzag (ZZ) nanotube. We notice that a ribbon with AC edges is terminated by
ZZ nanotubes and a ribbon with ZZ edges is terminated by AC nanotubes. 
Nanoribbons  terminated by AC nanotubes present interesting magnetic properties. 
By rolling the ZZ edges of a nanoribbon, two types of AC nanotubes can be formed, as shown in figure \ref{fig:zzacnt}.  
If the atoms at the nanoribbon ZZ edge  scroll and bind to the same sublattice
sites within the nanoribbon, the formed AC nanotube has mirror symmetry with 
respect to the nanoribbon plane; if the bonding sites belong to
opposite sublattice, there is no such kind of symmetry (compare figure \ref{fig:zzacnt}b to figure \ref{fig:zzacnt}a). 
We call these two cases \textit{%
symmetric} and \textit{asymmetric} which 
corresponds to \textit{armchair} and \textit{armchair-like} in Ref. \cite%
{Ivanovskaya}, respectively. The common point of these two cases is that
the sublattice symmetry is broken, because all the $sp^{3}$ hybridized carbon atoms
at the junction belong to one sublattice. Due to Lieb theorem \cite{Yazyev2010,KatslesonBook}, this gives the possibility of spin
polarization around the junctions. Since the theorem applies to the Hubbard model, accurate calculations for the real system are necessary to 
investigate this possibility. 

% \section{Edges Terminated by AC Nanotubes}

\begin{table}[tbp]
\caption{Total spin magnetization $S$ (in unit of $\protect\mu _{B}$ per unit cell) as a
function of the index of AC nanotube ($N,N$) and size of nanoribbon
$P$ (in unit of ZZ rows). The results in this table
are for samples obtained by rolling the edges of the same unrolled graphene ribbon with width of $40$ ZZ rows.}
\label{tab:spin}\centering%
\begin{tabular}{ccccccc}
\hline
% & symmetric &  & ~~~~~~~~~ &  & asymmetric &  \\ \hline
% N & P & S &  & N & P & S \\ \hline
% 9 & 4 & 0.000 &  & 9 & 4 & 1.499 \\ 
% 8 & 8 & 0.995 &  & 8 & 8 & 1.499 \\ 
% 7 & 12 & 1.371 &  & 7 & 12 & 1.499 \\ 
% 6 & 16 & 1.481 &  & 6 & 16 & 1.499 \\ 
% 5 & 20 & 1.500 &  & 5 & 20 & 1.500 \\ 
% 4 & 24 & 1.500 &  & 4 & 24 & 1.500 \\ 
% 3 & 28 & 1.494 &  & 3 & 28 & 1.500 \\ \hline

N & P & S (symmetric) & S (asymmetric)\\ \hline
9 & 4 & 0.000 & 1.499 \\ 
8 & 8 & 0.995 & 1.499 \\ 
7 & 12 & 1.371 & 1.499 \\ 
6 & 16 & 1.481 & 1.499 \\ 
5 & 20 & 1.500 & 1.500 \\ 
4 & 24 & 1.500 & 1.500 \\ 
3 & 28 & 1.494 & 1.500 \\ \hline

\end{tabular}%
\end{table}

In order to study the magnetic properties, we performed spin polarized DFT calculations by SIESTA\cite{SIESTA-1, SIESTA-2, SIESTA-3}. 
We used generalized gradient approximation with Perdew-Burke-Ernzerhof parametrization (GGA-PBE) \cite{GGA-PBE} and
a standard built-in double-$\zeta $ polarized (DZP) \cite{DZP-NAO} basis set
to perform geometry relaxation. 
We found that, in both symmetric and asymmetric cases, there is spin polarization near the ribbon-tube junction, i.e., near the $sp^{3}$ hybridized carbon atoms. 
Note that the bond distance of these four-fold coordinated atoms is $1.54$~\AA~ like in diamond.  
The spin polarization is mainly located in the nanoribbon for the symmetric
case and within the tube for the asymmetric case (see the isosurface plot
of the spin density together with its symbolic representation in figure \ref%
{fig:zzacnt}c-h). The up/down spins are distributed over the A/B sublattices
respectively. The label $m_{max}$  indicates the atom with the highest magnetic moment.

For the symmetric case, the value of the spin polarization
increases with increasing nanoribbon width, and saturates at
1.50 $\mu _{B}$ per unit cell, when the nanoribbon width is wider than 16 ZZ
rows; For the asymmetric case, the spin
polarization is always 1.50 $\mu _{B}$ per unit cell, irrespective of the 
nanoribbon width and nanotube radius (see Table \ref{tab:spin}).

For the asymmetric case the spins are located inside the two
nanotubes and therefore the exchange interaction between opposite edges is negligible.
For the symmetric case, the 
spins on the two edges are coupled  antiferromagnetically, similarly to  hydrogen terminated
graphene edges\cite{Yazyev,Bhandary}:  for the structure shown in figure \ref%
{fig:zzacnt}a, the energy of  antiparallel spin configurations is $22$ meV per unit cell (see figure \ref{fig:zzacnt}g) lower than for parallel configurations. 
This sizeable coupling across the ribbon makes the symmetric systems promising as spin valve devices\cite{Hill2006}. In figure \ref{app}a we show a configuration similar to that proposed for dumbbell graphene structures on the basis of the Hubbard Hamiltonian in the  mean field approximation\cite{MaZL2011}. A gate could be used to bring the system from the antiferromagnetically coupled state to the ferromagnetically coupled excited state, favouring spin transport from one nanotube to the other across the ribbon. 
Moreover, both for the symmetric and asymmetric case, the magnetic moments along the ribbon-tube junction are qualitatively similar to the case of ZZ edges of nanoribbons. Therefore, high magnetoresistance could be expected, as proposed in \cite{KimWY2008} for nanoribbons with ZZ edges, by applying magnetic fields of different sign at the ends of the nanoribbon.  A sketch of this device for our systems is shown in figure \ref{app}b.

\begin{figure}[t]
\begin{center}
\mbox{
\includegraphics[width=0.7\textwidth]{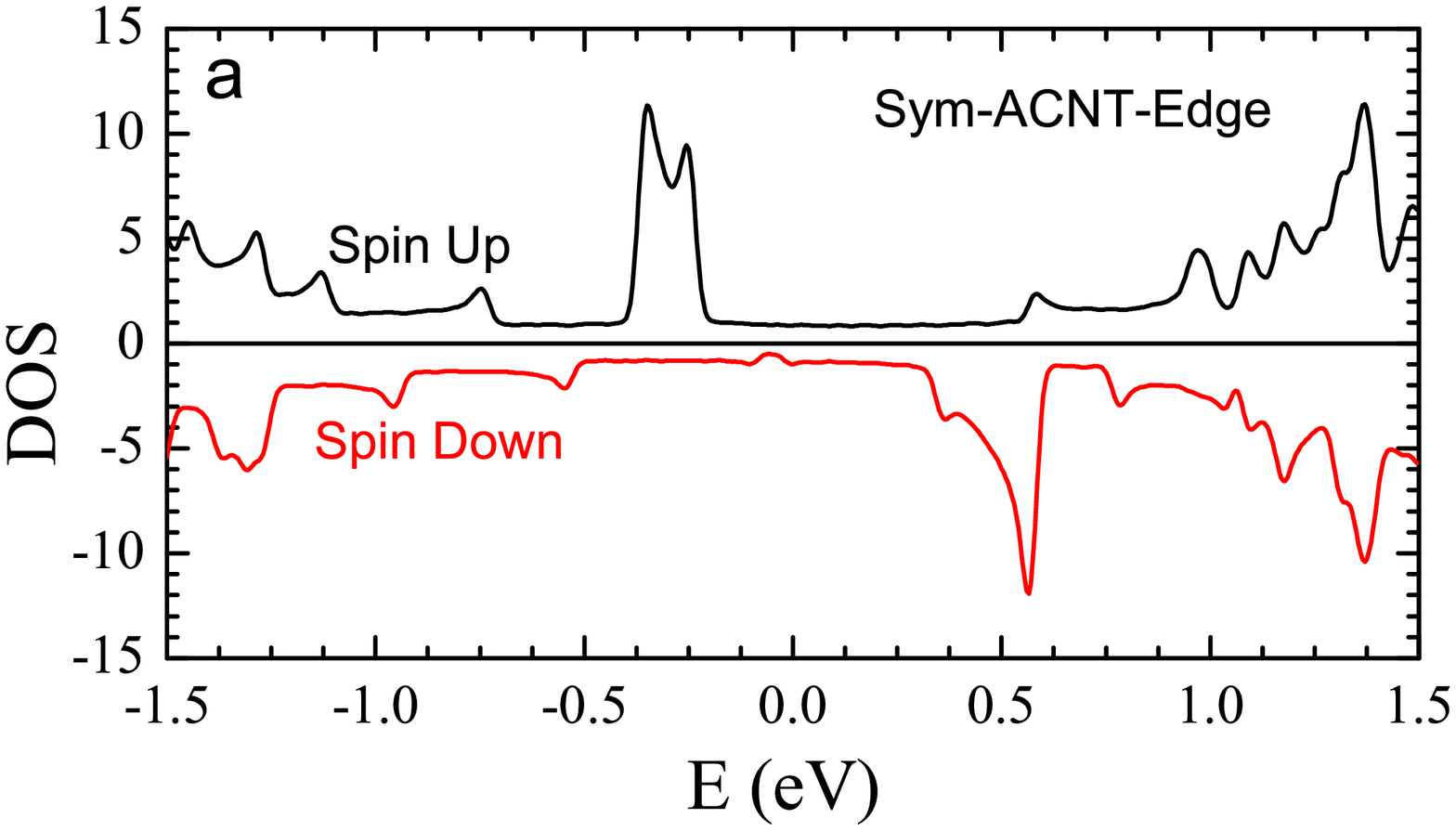}
}

\mbox{
\includegraphics[width=0.7\textwidth]{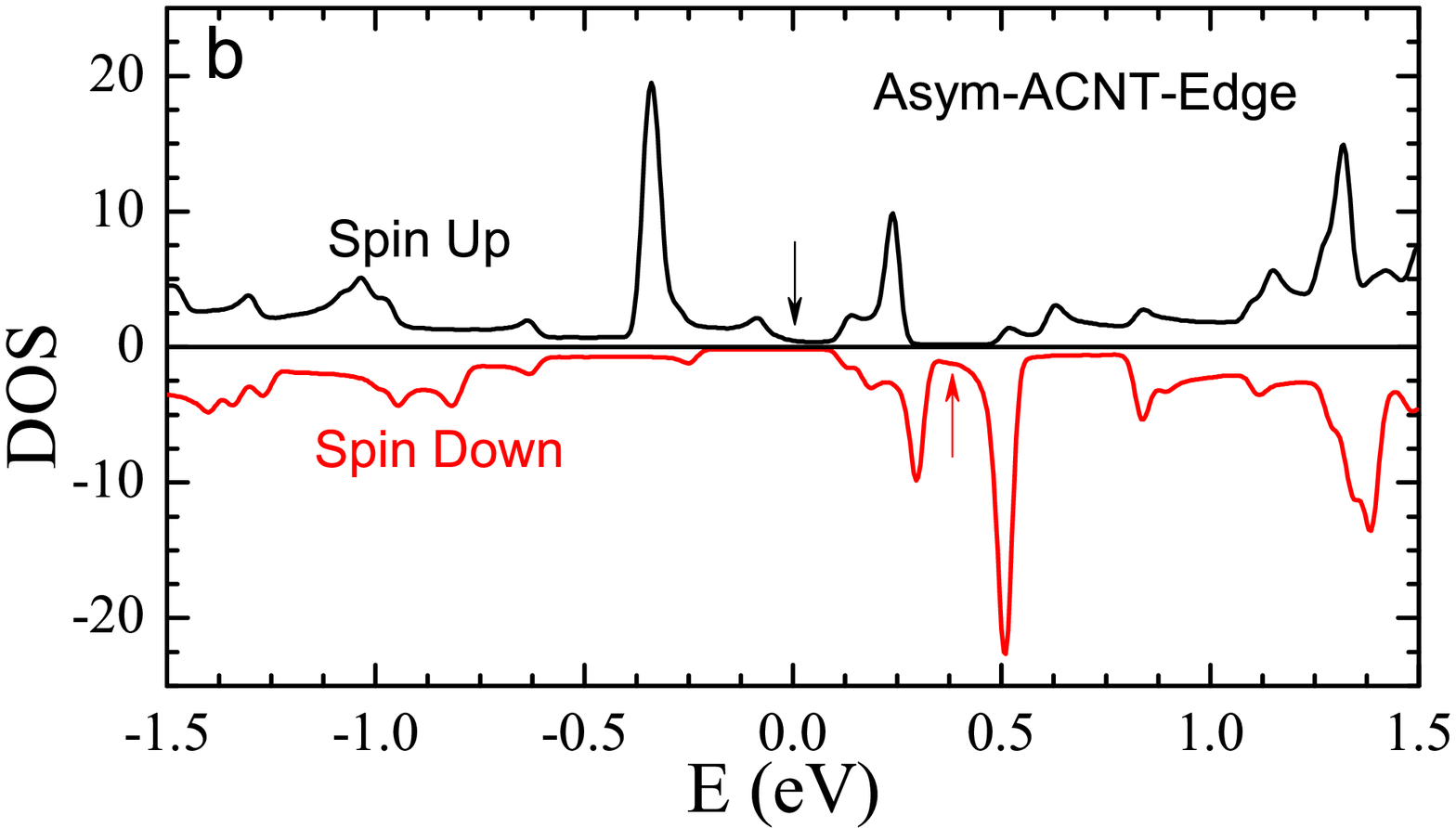}
} 
\end{center}
\caption{Density of states calculated by  DFT  for ZZ nanorribons 
with (a) symmetric and (b) asymmetric AC nanotube terminated edges ($N=5$ and $P=20$ for
both cases). The arrows in (b) indicate the half-metalic energy regions for either spin up or spin down.}
\label{ldosarmchair}
\end{figure}

The spin polarized density of states (DOS) reveals other features of interest for spintronics related to half-metallic character. 
In figure \ref{ldosarmchair}a and \ref{ldosarmchair}b we show the spin polarized DOS for the symmetric and asymmetric case respectively. We see that the symmetric case is metallic for both spins in the whole range of energy.  The asymmetric case, instead, is a half-metal near the Fermi energy $E_F$, namely it is metallic  for spin up and insulating for spin down.  The half-metallic character of our systems  provides opportunities as spin filters without the need of external electric fields\cite{SonYW2006}, magnetic fields\cite{Abanin2006}, ferromagnetic strips\cite{ZhangYT2010}, impurities\cite{Hod2007,ZhengXH2009,Soriano2010} or defects\cite{Martins2008,Lisenkov2012}. Furthermore, there is the opposite half-metallic character at higher energies. Around 0.4 eV, there is  insulating character for spin up and metallic character for spin down. As sketched in figure \ref{app}c, a gate along the ribbon could be used to switch  between these two half-metallic energy regions  and affect selectively  the spin transport.

% \section{Edges Terminated by ZZ Nanotubes}

\begin{figure}[tbp]
\centering
\mbox{
\includegraphics[width=0.75\textwidth]{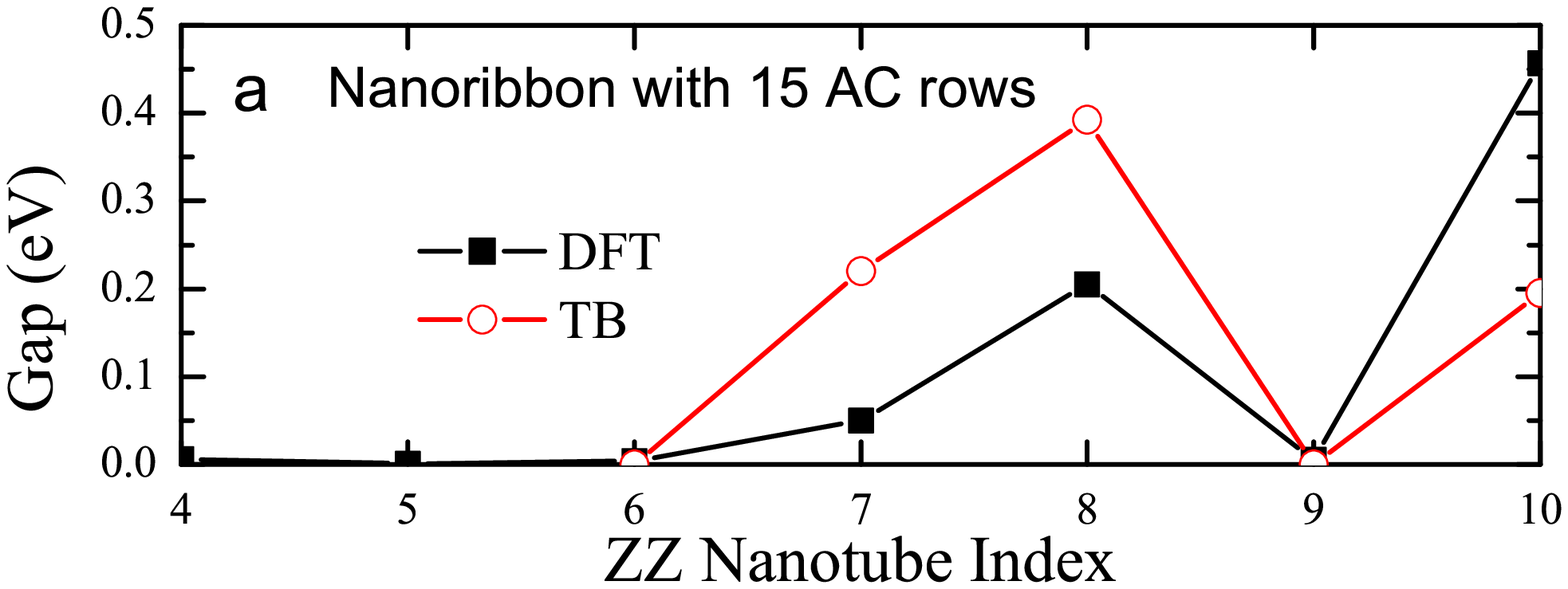}
} \mbox{
\includegraphics[width=0.75\textwidth]{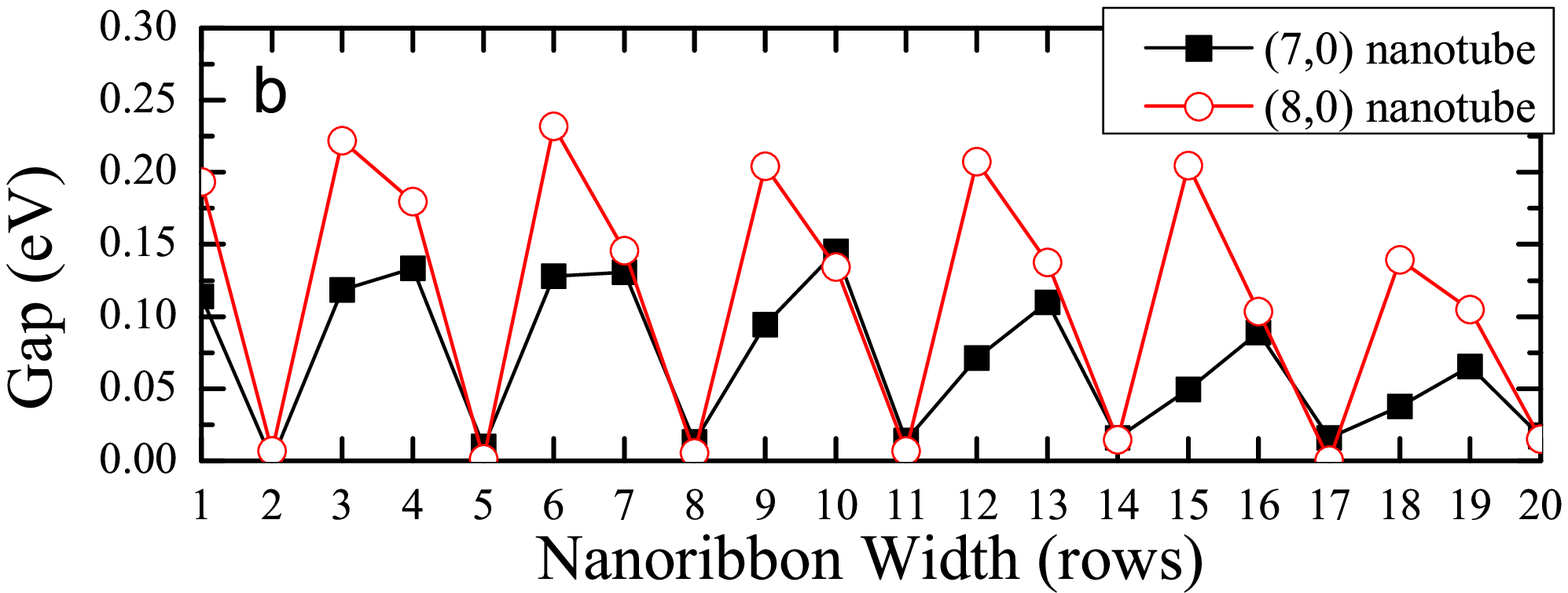}
}
\caption{Band gap size-dependence in ZZ nanotube terminated AC nanoribbons.
(a) Fixed nanoribbon with 15 AC rows for varying size of the terminating  ZZ nanotube.
(b) Fixed ZZ nanotube ($8,0$ and $7,0$) edges for varying
width of the AC nanoribbon. }
\label{fig:gaps}
\end{figure}

\begin{figure}[t]
\begin{center}
\mbox{
\includegraphics[width=0.7\textwidth]{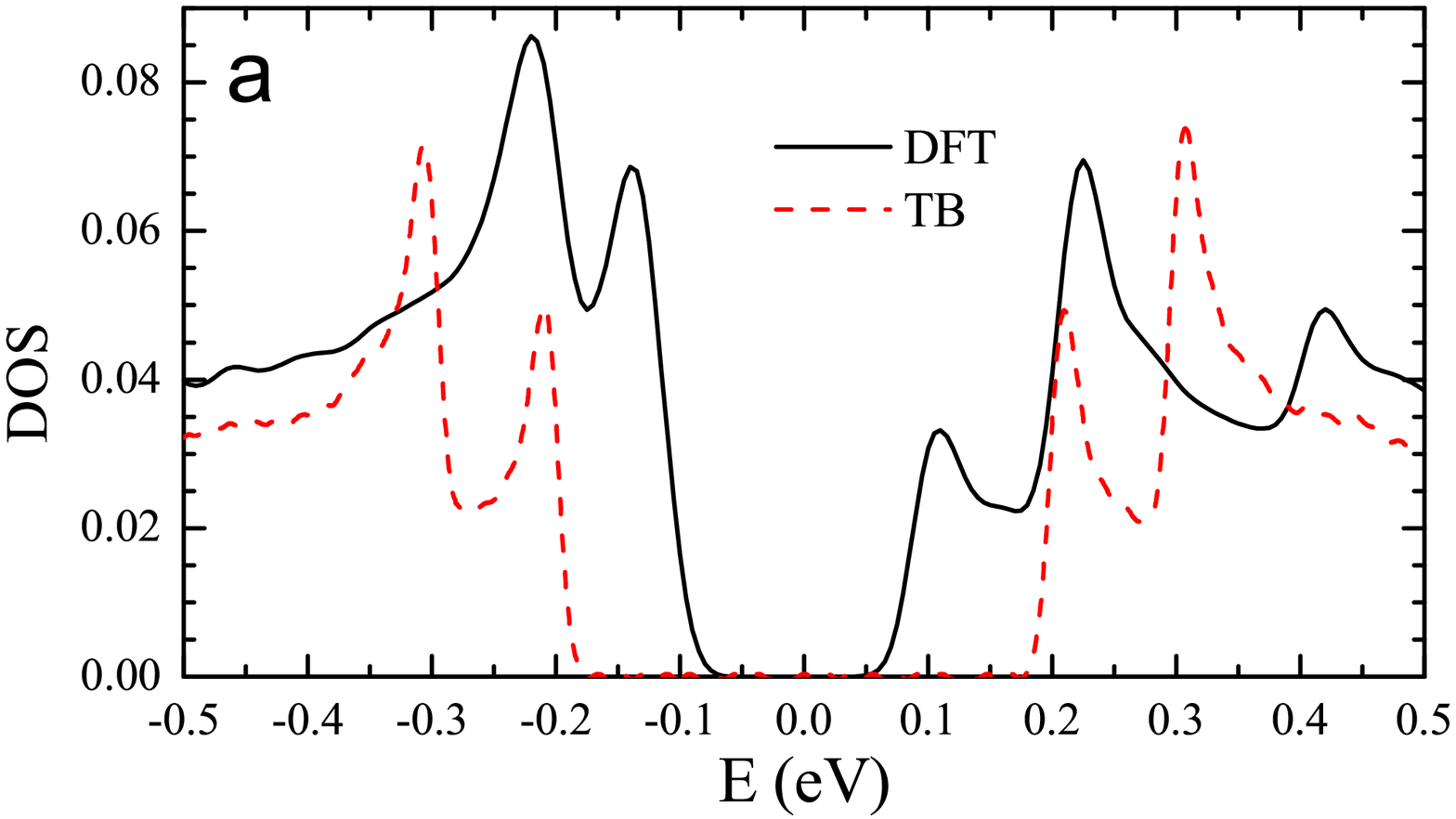}
} \mbox{
\includegraphics[width=0.7\textwidth]{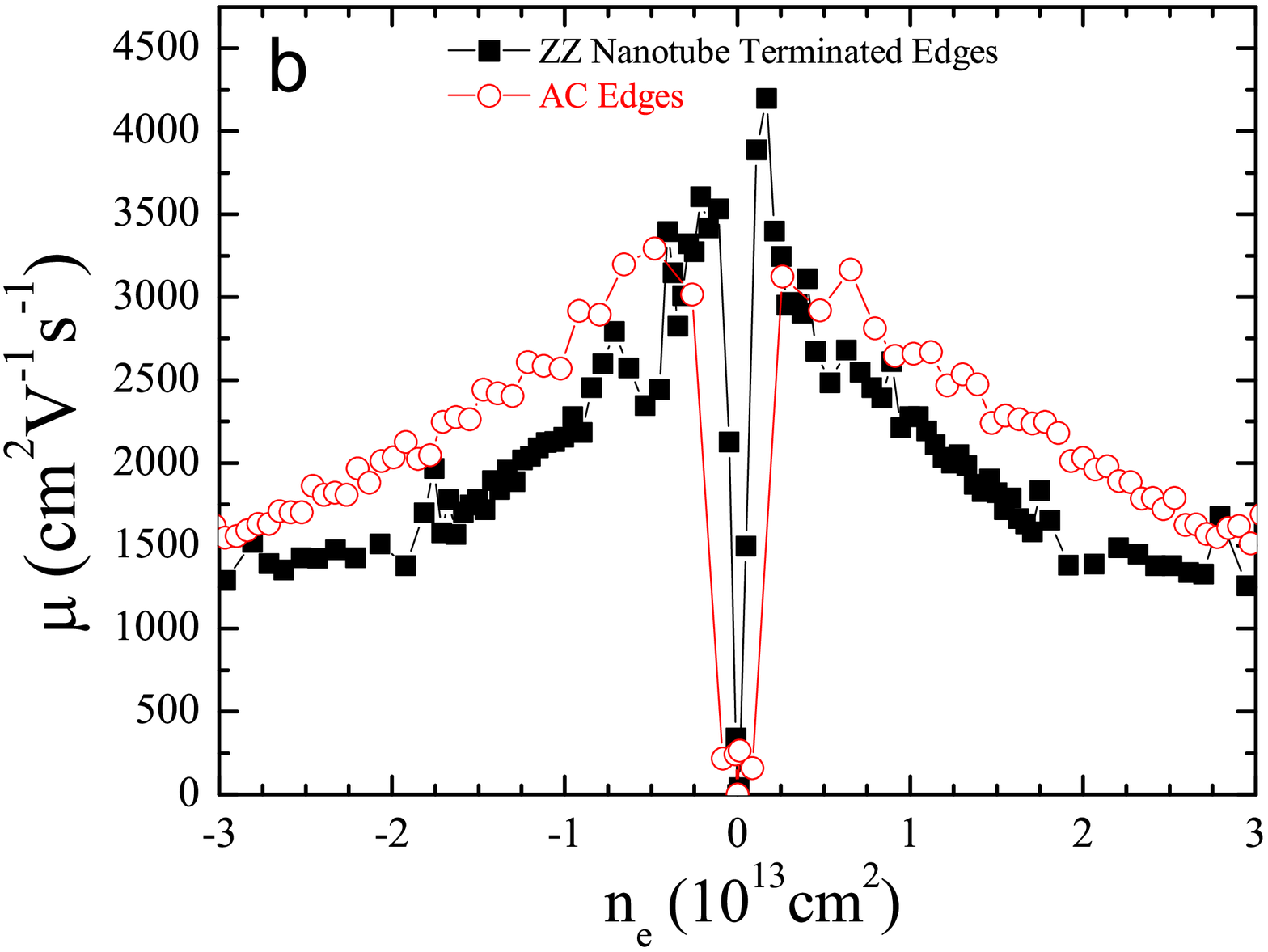}
}
\end{center}
\caption{(a) Comparison of DOS of a  nanoribbon with 15 AC rows terminated by
 (8,0) ZZ nanotubes calculated by DFT and TB.
(b) Comparison of the mobility of a nanoribbon with 15 AC rows, terminated either by (8,0) ZZ nanotubes or by AC edges in TB. The charge
density $n_{e}$ is obtained from the density of state $\protect\rho $ in
panel (a) by $n_{e}\left( E\right) =\protect\int_{0}^{E}\protect\rho %
\left( \protect\epsilon \right) d\protect\epsilon $. The sample
contains $600000\times 47$ carbon atoms for the case with ZZ nanotube
terminated edges, and $2000000\times 15$ for the case with AC edges.}
\label{dczigzag}
\end{figure}

We come now to the case with ZZ nanotubes shown in figure \ref{fig:zzacnt}i-j. For this case,  there is only one type of ribbon-tube junction that preserves sublattice symmetry 
implying that there is no magnetization nor midgap states\cite{Yazyev2010,KatslesonBook}. The electronic
structure and transport properties, however,  strongly depend on the AC ribbon width 
and on the ZZ tube radius. In TB models, a AC nanoribbon  is metallic
 if the number of AC rows is equal to $3l+2$, where $l$ is a positive
integer, and semiconducting otherwise\cite{Son2006,Ezawa2006}. Furthermore, ZZ
nanotubes are metallic for index equal to  $3l$\cite{Dubois}. In more
general models, the properties of AC nanoribbons and ZZ nanotubes
may differ from the ones predicted by TB, due to possible  self
passivation of the edges for nanoribbons and for the $\sigma -\pi $ band mixing for small
nanotubes\cite{Dubois}. By using DFT calculations, we found that our joined system
becomes a semiconductor with a gap of the order of few hundreds meV
if both nanoribbon and nanotube are semiconducting. The energy
gap as a function of geometry is shown in figure \ref{fig:gaps}. The size of the
nanotube has to be large enough for the opening of a band gap (figure %
\ref{fig:gaps}a). For the joined system with semiconducting ZZ
nanotubes, there is a clear periodicity (3 ZZ rows) in the
dependence of the energy gap on the nanoribbon width (figure \ref%
{fig:gaps}b). For the studied cases, the value of the gap varies between 30 and
600 meV. 

Since we want to calculate the transport properties by means of a simpler model, suitable for large samples, 
we have also calculated the energy structure of our system by  $\pi $-band TB calculations where we consider only the nearest-neighbor
hopping $t=2.7$ eV between the carbon atoms\cite{WK10,YRK10} for all bonds. The comparison between the gaps calculated by TB and DFT shown in figure \ref{fig:gaps}a gives the same periodicity and a qualitative agreement. For the case of a AC nanoribbon with 15 rows terminated by (8,0) ZZ nanotubes we compare in figure \ref{dczigzag}a the TB and DFT DOS, which again are in qualitative agreement as to support the validity of the transport calculations we show next. 
The electronic transport properties of a semiconducting 
nanoribbon with or without nanotube terminated edges are obtained by
using  large scale TB simulations with about thirty million carbon atoms%
\cite{WK10,YRK10}. In figure \ref{dczigzag}b we show that the electronic mobility parallel to the edges  as a function of charge density are
quite similar in these two cases. The mobility of the joined system 
is about $2250$  $cm^{2}V^{-1}s^{-1}$ at charge density $n_{e}\sim 10^{13}cm^{2}$, which is only slightly smaller than the one of the AC nanoribbon 
at the same charge concentration.

% \section{Conclusion}

In summary, we have studied the electronic and magnetic properties of graphene nanoribbons
with three types of nanotube terminated edges. The spin magnetization is
found to be 1.5 $\mu _{B}$ per unit cell in the ground state of both
symmetric and asymmetric AC nanotube terminated edges. For symmetric
AC nanotube terminated edges, the spin density is located in the ribbon whereas, for the asymmetric case, it is located within the tube.
In the ZZ nanotube terminated edges, there is a band gap opening of the
order of few hundreds meV, if the constituent tube and 
nanoribbon are both semiconducting. The conductivity and mobility
 in the presence of ZZ nanotube terminated edges is 
comparable to the one of the AC nanoribbon itself.

Our calculations suggest that these systems are not only advantageous because the edges
 are protected against any kind of chemically induced disorder but also because, by tailoring the  ribbon/tube structure, they offer a wealth of possible applications for transport and spintronics. 

\section*{Acknowledgments}
The support by the Stichting Fundamenteel Onderzoek der Materie (FOM) and
the Netherlands National Computing Facilities foundation (NCF) are
acknowledged. We thank the EU-India FP-7 collaboration under MONAMI and the
grant CONSOLIDER CSD2007-00010.

\section*{References}
% \begin{thebibliography}{10}
\bibliographystyle{iopart-num}
\bibliography{BibliogrGrafeno}

\end{document}